\documentclass[pra, twocolumn]{revtex4-1}
\usepackage{graphicx}
\usepackage{amstext}
\usepackage{amsmath}
\usepackage{color}
\usepackage{soul}

\newcommand{\De}[1]{\left[ #1 \right]}

\newcommand{\ket}[1]{\left| #1 \right\rangle}
\newcommand{\bra}[1]{\left\langle #1 \right|}

\begin{document}

\title{Double-slit implementation of minimal Deutsch algorithm}
\author{B. Marques$^{1}$, M. R. Barros$^{1}$, W. M. Pimenta$^{1}$, M. A. D. Carvalho$^{1}$,
J. Ferraz$^{1}$, R. C. Drumond$^{1,2,3}$, M. Terra Cunha$^{3}$, and S.
P\'adua$^{1}$}

\affiliation{$^{1}$Departamento de F\'isica, Universidade Federal de Minas Gerais,caixa postal 702,
30123-970, Belo Horizonte, MG - Brazil}
\affiliation{$^{2}$Instituto Nacional de Matem\'atica Pura e Aplicada IMPA Estrada Dona Castorina,
110 Jardim Bot\^anico 22460-320, Rio de Janeiro, RJ, Brazil}
\affiliation{$^{3}$Departamento de Matem\'atica, Universidade Federal de Minas Gerais, caixa postal
702, 30161-970, Belo Horizonte, MG - Brazil}
\begin{abstract}

We report an experimental implementation of the minimal Deutsch algorithm
in an optical setting. In this version, a redundancy is removed from the most
famous form of the algorithm. The original version involves manipulation of two
qubits, while in its minimal version, only one qubit is used. Our qubit is
encoded in the transversal spatial modes of a spontaneous parametric down-converted signal photon,
with the aid of a double slit, with the idler photon
playing a crucial role in creating a heralded single photon source. A spatial
light modulator (SLM) is programmed to physically generate one-bit functions
necessary to implement the algorithm's minimal version, which shows that the SLM
can be used in future implementations of quantum protocols.

\end{abstract}
\maketitle

\section{Introduction}

Quantum computation has emerged in the past decades as a potentially powerful tool to solve problems
more efficiently than its classical counterpart. One simple example is determining whether a coin is
fair (heads on one side, tails on the other) or fake (heads or tails on both sides). This is one
version of \emph{the Deutsch problem} \cite{DEUTSCH}, who himself showed that, when exploring
quantum state superposition, only one examination step is necessary, while classically the solution
requires individual examination of both sides. Although the algorithm originally proposed by Deutsch
involves manipulation of two-qubit states, there is also a ``minimal'' version of it, in the sense
that just one qubit is manipulated \cite{BETTER}. In this sense, minimal Deutsch algorithm can be
considered the most basic and simple quantum computation.

Many physical systems have shown to be useful for implementing
quantum computation, such as nuclear magnetic resonance \cite{NMR1,NMR2}, trapped ions \cite{ION1},
optical
cavities \cite{CAV}, Josephson junctions \cite{JOSEPH}, and
photons \cite{PHOTON1,PHOTON2}. In particular, the original Deutsch
algorithm and its generalization, the Deutsch-Jozsa
algorithm \cite{DEUTSCH1992rapid}, were implemented using
photons \cite{PHOTON2,OLIVEIRA2005implementing},
and its minimal version was also implemented on quantum
dots \cite{BIANUCCI2004single}.

Spontaneous parametric down conversion (SPDC) \cite{SPDC} is a
natural source of correlated photon pairs, with the additional
advantage of having many degrees of freedom that can be considered quantum systems, such as
polarization \cite{SPDC,POLARIZATION},
transversal \cite{MOMENTA},
longitudinal \cite{LONGITUDINAL1,LONGITUDINAL2}, and orbital
angular \cite{ORBITAL1,ORBITAL2} momenta. On the other hand, a
spatial light modulator (SLM) can be used to perform state
control \cite{SLM1}. It has been used for tomographing
polarization \cite{SLM2} and transverse momenta \cite{SLM3} states, for measuring Bell inequality
violations in orbital momenta of
SPDC photon pairs \cite{SLM4}, and optical quantum algorithm simulation \cite{SIMULATION}.

In this work we report an experimental implementation of the
minimal Deutsch algorithm in an optical setting. We use a double slit to encode logical qubits (in
the sense of the $\left\{\left|0\right\rangle, \left|1\right\rangle\right\}$ logical base of a qubit
space) in the transversal spatial modes of photons generated in a SPDC process \cite{MOMENTA}. A SLM
is employed as a fundamental part of our experimental setup. It has the function of simulating the
``fair coin'' or the ``fake coin'' in the optical setup. Before really applying the quantum
algorithm, we need a calibration process, which can also be understood as a proof of principle of
the algorithm, since one uses many caries of the oracle. After such calibration, the apparatus is
ready for running the real Deutsch algorithm: With only one ``examination step'', answer which type
of ``coin'' we have, with probability larger than 1/2. The paper is organized as follows: In
Sec. II we review the theoretical description of the minimal Deutsch algorithm. The experimental
setup is presented in Sec. III. The results are shown in Sec. IV. Discussion is made
in Sec. V and conclusions are outlined in Sec. VI.

\section{Minimal Deutsch Algorithm}

Quantum parallelism allows quantum systems to evaluate a function
$f(x)$ for many different values of $x$ simultaneously.
The Deutsch algorithm is a good example of how to explore quantum
parallelism to answer a classical question; explicitly, to solve the Deutsch
problem evaluating the function only once.

Consider an \emph{oracle} that can answer one-bit questions with one bit answers described by a
deterministic function
$f:\{0,1\}\rightarrow\{0,1\}$. This function is called balanced if
$f(0)\neq f(1)$, otherwise the function is
constant [$f(0)=f(1)$]. The Deutsch problem consists in
determining whether a given unknown function is balanced or constant. 
For a classical algorithm to answer that with certainty it requires 
the oracle to be asked twice, that is, asking the value of $f$ on $0$
\emph{and} $1$. On the other hand, the Deutsch algorithm
requires only a single query, using quantum parallelism, to reduce the
minimal resource required.

In the minimal version \cite{BETTER},
the oracle's behavior is encoded in a unitary
operation $U_{f}$ to be applied on a well-chosen  input state, depending on the function $f$. For
the computational basis one has
 $U_{f}\ket x=(-1)^{f(x)}\ket x$ as output, for $x=0,1$.
 Now, if the superposition state $\frac{1}{\sqrt{2}}(\ket0+\ket1)$ is used
as input, both questions are asked at the same time, with the output state being:
\begin{equation}
\ket\psi=U_{f}\frac{\ket0+\ket1}{\sqrt{2}}=\frac{(-1)^{f(0)}\ket0+(-1)^{f(1)}\ket1}{\sqrt{2}}.
\end{equation}
Disregarding global phases, one has:
\begin{equation}
\ket\psi=\left\{ \begin{array}{rc}
\ket{+} = \frac{1}{\sqrt{2}}(\ket0+\ket1), \qquad \text{if} \quad f(0)=f(1),\\
\ket{-} = \frac{1}{\sqrt{2}}(\ket0-\ket1), \qquad \text{if} \quad
f(0)\neq f(1). \end{array}\right.
\end{equation}
Note that the first answer is orthogonal to the second, so we can
make a projective measurement in the basis $\{\ket+,\ket-\}$ and find out if the function is
balanced ($\ket-$) or constant ($\ket+$).
The unitary operations are implemented by the oracle. 
Four possible maps $U_{ij}$ are generated and we label them with the 
values of $i=f(0)$ and $j=f(1)$:
\begin{eqnarray}
\label{evol}
  U_{00} &=& \De{\begin{array}{cc} 1&0\\ 0& 1\end{array}}  \qquad\qquad f(0)=f(1)=0,\nonumber \\
  U_{01} &=& \De{\begin{array}{cc} 1&0\\ 0& -1\end{array}} \qquad\quad f(0)=0 \neq f(1)=1,\nonumber
\\
  U_{10} &=& \De{\begin{array}{cc} -1&0\\ 0& 1\end{array}} \qquad\quad f(0)=1 \neq  f(1)=0,\nonumber
\\
  U_{11} &=& \De{\begin{array}{cc} -1&0\\ 0& -1\end{array}}\qquad\quad f(0)=f(1)=1.
\end{eqnarray}

\section{Experimental Setup}

The scheme of the experimental setup is shown in Fig.1. A 50 mW He-Cd laser operating at $\lambda$ =
325 nm is used to pump a 2-mm-thick lithium iodate crystal and generate, by type I SPDC, degenerate
non-collinear photon pairs. Signal and idler ($\lambda_{s}=\lambda_{i}$ = 650 nm) beams pass through
a $\lambda/2$ plate (half-wave plate), before they cross a double-slit placed at a distance of 250
mm from the crystal. The double-slit plane ($xy$ plane) is aligned perpendicular to the plane
defined by the pump laser and the down-converted beams ($yz$ plane), with the small dimension of the
slits parallel to the $x$ direction. The slits are $2a = 100$ $\mu$m wide and have a separation of
$2d = 250$ $\mu$m. The lens $L_{1}$ is used to generate photon pairs in entangled transversal path
states~\cite{LEONARDO}. A natural question is why to use down-converted biphotons for implementing
the one-qubit Deutsch algorithm. In our case, one can consider the selective detection of the idler
photon as part of the heralded 
preparation of the signal photon state. 

\begin{figure}[ht!]
\hspace{0.1cm}
\begin{center}
\includegraphics[width=9cm]{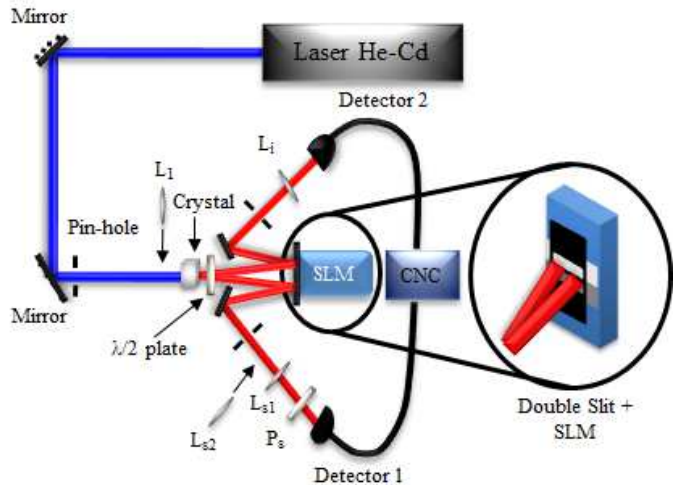}
\end{center}
\caption{(Color online) Experimental setup scheme for minimal Deutsch algorithm implementation. The
$L_{1}$ lens
focuses the pump beam in the double-slit plane; lenses $L_{s1}$ and $L_{i}$ are used to detect the
signal and idler beams at Fourier plane, while the $L_{s2}$ lens is used to project the double slit
images in the detector. A half-wave plate is placed right after the crystal and a polarizer $P_{s}$
is positioned in front of detector 1. CNC denotes coincidence counter and SLM denots spatial
light modulator.}
\label{fig:Fig1}
\end{figure}

Our experimental setup is arranged in a way that a photon passing through the inferior slit of the
double slit corresponds to
state $\ket0$, while a photon that passes through the superior slit corresponds to state $\ket1$.
The SLM after the double slit, together with the $\lambda/2$ plate and the polarizer $P_s$, works as
the quantum oracle. The map of the oracle function is constant ($U_{00}$ or $U_{11}$) if the phases
added by the SLM are equal, or balanced ($U_{01}$ or $U_{10}$) if the phase difference is $\pi$.
Once the pump beam is focused at the double slit plane, a Bell state $\ket{\psi_+}$ is created by
the twin photons in the slit path states \cite{LEONARDO,EMARANHANDO}. A dichroic mirror placed just
after the crystal removes the pump beam and transmits signal and idler beams. The trigger photon
(idler) also passes through a double slit and is reflected by the SLM, but without the polarizer at
its path, no phase change or amplitude variation in its state occurs due to the manner in which the
SLM
works. In Fig.~\ref{fig:Fig2} and Fig.~\ref{fig:Fig3}, we show, experimentally, that we are able to
introduce spatial phase 
changes at the signal-photon path state while preserving the state amplitudes by using the SLM.
However, there are many other maps that the SLM can implement in a double slit-qubit. The SLM
maps can
be described by:
\begin{equation}
U_{SLM} = \De{\begin{array}{cc}A_{0}e^{i\phi_{0}}&0 \\ 0&A_{1}e^{i\phi_{1}}\end{array}},
\end{equation}
where $A_{k}$ is the attenuation  and $\phi_{k}$ is the phase applied in the photon state
$\ket{k}$. The SLM maps in this kind of setup are diagonal because it cannot exchange photon
population between slits, i.e., an operation like $\ket{0(1)}\mapsto A_{0}\ket{0}+A_{1}\ket{1}$
cannot be done, if $A_k \neq 0$. Instances of these SLM maps were implemented in Refs.~\cite{SLM2,
SLM3}, while a general one can be made through a calibration described by Moreno \textit{et al.}
\cite{MORENO}.

Single slits with 100 $\mu$m width are placed in front of each detectors. Their planes ($xy_{i,s}$
planes) are aligned perpendicular to the propagation direction of
the idler and signal beams ($z_{i,s}$ direction), respectively.  The
small dimension of each slit is parallel to the corresponding
$x$ direction. The SLM used is a Holoeye Photonics LC-R 2500, which
has a $1024 \times 768$ pixel resolution (each pixel consists of a $19 \times 19$
$\mu$m square) and it is controlled by a computer. Signal and idler
beams are focused on the detectors with a microscope objective
lens (not shown in Fig.~\ref{fig:Fig1}). Two interference filters,
centered at 650 nm and 10 nm FWHM bandwidth, are kept before the objective
lenses.
Pulses from the detectors are sent to a photon-counter and a
coincidence detection setup with a 5.0 ns resolving time.

\section{Experimental Results}

First of all, we must be able to implement the maps $U_{ij}$ ($i,j\in\{0,1\}$) of Eq.~\eqref{evol}.
It was shown \cite{MORENO} that a SLM plus the input and output
polarizers can be properly calibrated to obtain this goal ($A_{0}=A_{1}$ and $\phi_{k}=0,\pi$). The
liquid crystal display of the SLM is divided in two regions, each
region adding a phase to the photon path state, $\ket0$ or $\ket1$, defined by the slits (inset of
Fig.~\ref{fig:Fig1}). 
A SLM gray level is associated with each region of the display. Pre-determined gray
levels, along with correct half-wave plate and polarizer
angles, introduce a relative phase between the photon path states
without relative amplitude attenuation. The evolution maps $U_{ij}$ are
implemented when the correct phase differences (0 or $\pi$) with no
amplitude attenuation are introduced in the photon path states by the SLM.

\begin{figure}[ht!]
     \begin{center}
        \includegraphics[width=9cm]{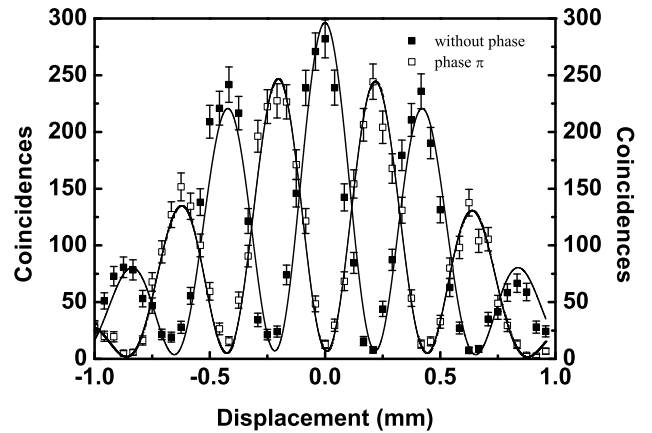}
        \end{center}
        \caption{Coincidence double-slit interference patterns. Closed squares show the interference
pattern when the SLM gray levels are the same in both slit aperture directions that define the
signal photon paths states. Open squares show the interference pattern when there are different gray
levels producing a relative phase of $\pi$, between the signal states. Idler path phase, in both
measurements, is not changed by the SLM. The idler detector is kept fixed at x$_{i}$ = 0, while
signal detector is scanned in steps of 40 $\mu$m and the detection time is 60 seconds. Lenses
L$_{i}$ and L$_{s1}$ were used.}
\label{fig:Fig2}
\end{figure}

Figure \ref{fig:Fig2} shows two double-slit interference patterns, measured in coincidence counts.
Both are obtained by scanning the signal beam detector, with steps of 40 $\mu$m, while maintaining
the idler beam detector fixed in the position that corresponds to an interference pattern maximum
($x_i$ = 0). In the closed squares pattern, we have used the same SLM gray level for the $\ket0$ and
$\ket1$ signal photon path states. But in the open squares pattern we have inserted a relative phase
between the photon states, through an appropriate choice of SLM gray levels for each one. Fitting
the two interference graphs, we measure a relative phase of $\Delta\phi=3.25\pm0.03$. To obtain the
interference pattern in both detectors, we have used the L$_{i}$ and L$_{s1}$ lenses in our
experimental configuration. It is important to note that the phase in the idler photon path states
is not affected by the SLM gray level, because there is no polarizer in front of this
detector~\cite{MORENO}.

\begin{figure}[htpb]
    \begin{center}
        \includegraphics[width=9cm]{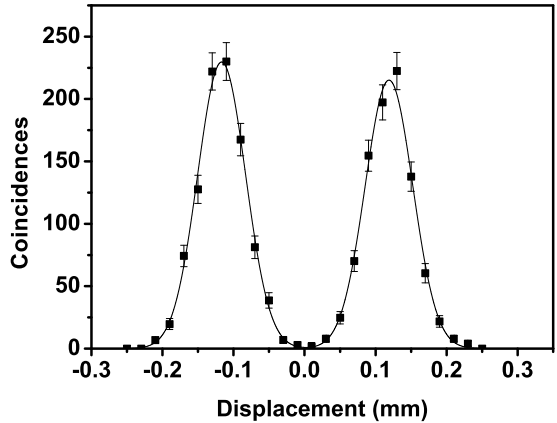}
        \end{center}
        \caption{Signal double-slit image, measured in coincidence counts. The result is recorded
with the idler detector fixed at x$_{i} = 0$, while signal detector is scanned in the x direction,
when the same gray level of open squares pattern used in Fig.~\ref{fig:Fig2} is applied and the
detection time is 20 seconds. Lenses L$_{i}$ (the same lens used in Fig.~\ref{fig:Fig2}) and
L$_{s2}$ were used. Signal detector is placed at the image plane while idler detector is at the
Fourier plane.}
\label{fig:Fig3}
\end{figure}

In Fig.~\ref{fig:Fig3} we have the coincidence double-slit image, for the signal beam. This result
is obtained, when the L$_{s2}$ lens is used in the signal beam and the L$_{i}$ is used in the idler
beam. 
Here the experimental setup is such that the peak in the x$_{s}$ displacement negative region is
associated with the inferior slit, \textit{i.e.}, with the $\ket0$ signal photonic state, while the
peak in the positive displacement region is associated with the $\ket1$ photonic state. Once again
we kept the idler detector fixed at $x_{i} = 0$, \textit{i.e.}, its interference pattern maximum and
scan the signal beam detector, with steps of 20$\mu$m and acquisition time of 20s. Signal detector
is placed at the image plane while idler detector is at the Fourier plane. The SLM gray levels were
the same as the ones used to obtain the open squares interference pattern shown in
Fig.~\ref{fig:Fig2}. We can infer that the SLM gray levels used to obtain phase $\pi$ do not
attenuate the state amplitude by calculating the areas under the peaks corresponding to each state.
In the curve shown, we have an area of $\left(77\pm2\right)$ arb.~units, for the $\ket0$ signal
photon
state and $\left(72\pm2\right)$ arb.~units, for
the $\ket1$ state. Therefore, with the curves shown in the open squares of Fig.~\ref{fig:Fig2} and
 Fig.~\ref{fig:Fig3}, we can implement the $U_{01}$ map. To implement each $U_{ij}$ a proper gray
level was chosen in each slit, corresponding to add a $\pi$ phase for the value $1$ and no extra
phase for the value $0$. 

To implement the minimal Deutsch algorithm we must create the state $\ket+$ for the signal
photon
and, after the oracle, measure it in the base $\{\ket+,\ket-\}$.
By detecting at the Fourier plane and at the origin of the interference pattern, we are able to
implement experimentally the detection projector $\ket{+}\bra{+}$ \cite{EMARANHANDO}. Using SPDC, a
double-slit, and by focusing the pump beam at the double-slit's plane, we prepare the Bell state
$\ket{\psi_{s,i}}=\ket{\psi^{+}}=\frac{1}{\sqrt{2}}(\ket{0_{s}1_{i}}+\ket{1_{s}0_{i}})=\frac{1}{
\sqrt{2}}(\ket{+_{s}+_{i}}-\ket{-_{s}-_{i}})$\cite{LEONARDO}.
If we detect the idler photon using the projection operator $\ket+\bra+$ we
project the signal photon state in the state we need. 
Hence, in our setting, an oracle query corresponds to a detection of the idler photon at the center
of the pattern, being the instance where the logical qubit, the signal photon, is prepared in the
appropriate state. The oracle's answer is then provided by the detection or no-detection of the
signal photon. Note, moreover, that the idler measurement, and hence logical qubit state
preparation,
is done after one of the maps $U_{ij}$ is applied on the qubit. But since these operations commute,
this does not affect the final statistics.

The experimental results for all map possibilities are shown in Table \ref{table1}. The measured
photon coincidences were obtained
with signal and idler detectors kept fixed at $x_{s} = 0$, and $x_{i} = 0$ (center of the
interference pattern), respectively,
at the Fourier plane.  Measurements were taken in $1 000$ s. The data show a clear difference of
behavior between constant and balanced functions.

\begin{table}
\caption{\label{table1} Experimental results for all maps possibilities.}
\begin{center}
\begin{tabular}{|c|c||c|}
\hline
$f(0)$ & $f(1)$ & Coincidence\\
\hline\hline
0 & 0 & 5218 $\pm$ 72\\
\hline
0 & 1 & 450 $\pm$ 21\\
\hline
1 & 0 & 427 $\pm$ 20\\
\hline
1 & 1 & 5399 $\pm$ 73\\
\hline
\end{tabular}
\end{center}
\end{table}

\section{Discussion}

In the previous section we presented the result of the algorithm using
many oracle queries, summarized in Table \ref{table1}, which allows, as expected,
perfect discrimination between constant and balanced functions (high and low coincidence counts,
respectively). The Deutsch algorithm is, however, about the optimization of the Deutsch problem with
regard to the number of queries to the oracle; and here we discuss what our experimental setup tells
us when only one such query is allowed.

However, the meaning of this implementation regarding individual
events is more subtle, due, among other reasons, to the fact that an idler photon detection
at
position $x$ corresponds to the preparation of a state
$\frac{1}{\sqrt{2}}(\ket{0}+e^{i\phi(x)}\ket{1})$ for the signal photon, where $\phi(x)$ is
a function depending on the detector position and other features of
the setting (see Refs.\cite{LEONARDO,EMARANHANDO}). Strictly speaking, the
measurement is not of the von Neumman type, on the states
$\{\ket{+},\ket{-}\}$, but a positive operator valued measurement (POVM).
This means that an infinitesimal detector in $x=0$ measures an operator proportional to
$\ket{+}\bra{+}$, 
while a finite detector with opening $d$ (that is, the width of the slit placed in front the
detector) measures a positive operator that 
is a weighted sum of all projectors from $x=-\frac{d}{2}$ to $x=\frac{d}{2}$. Moreover, a signal
no-detection event does not necessarily correspond  to the state $\ket-$. Nevertheless, we can use
the results shown in Fig.~\ref{fig:Fig2} to predict the chances of having a detection (and hence,
also no detection) event when the function is constant or balanced, by computing the area under the
corresponding (normalized) curve over an interval centered at the origin with the same width as the
detector.

From the graphs themselves, and also from Table \ref{table1}, we see that the chances of detecting a
photon when the function is constant is much higher then when it is balanced, so indeed a single
oracle query gives us some information about the function. But, in practice, none of the events tell
us \emph{definitely} which type of function we have. A detection can also be associated with a
balanced function, although it is rare, due to detector width. A no-detection
event, on the other hand, might be related to one of three distinct situations: the
function is balanced so, as we want, there is a very low probability of detecting a
photon at the origin; or the function is constant, but the detector
failed; the function is constant, but the photon hit the Fourier plane at
another point away from the detector. Due to this
third situation, even in a perfect experimental setting a no-detection event can also correspond to
a constant
function.

To understand better the quantum advantage within this implementation we can consider a
scenario where one of these functions is given to us, with equal
probability, and we must bet on constant or balanced with only one oracle
query. Classically, we cannot do better than a fifty-fifty guess,
but with this implementation we can. Indeed, from the discussion in the last paragraph, we already
see that a detection is more likely to be associated with a constant function, while no detection is
more likely to be associated with a balanced one. 

To see this in a quantitative manner, let $S$ be the event where the function's type
is correctly guessed. We can write the probability for this event as:
\begin{equation}
 P(S)=\sum_{i,j=0,1}P(f=ij)P(S|f=ij),
\end{equation}
where $P(f=ij)=1/4$ is the probability of having the function $f(0)=i,f(1)=j$ while $P(S|f=ij)$
is the
probability of success given that function. We denote by $p_{ij}$ the probability for a photon to
hit the Fourier plane at a point covered by the detector, given that the function implemented is
$f=ij$. That is, $p_{ij}$ is just the area under the normalized curve of Fig.~\ref{fig:Fig2}
corresponding to the function $ij$, in an interval around the origin with the detector's width.

Now, if the function
is the constant $00$, we succeed in our guess if we detect a photon. This will take place with
probability $\eta p_{00}$, where $\eta$ is the
detector efficiency. Similarly, for $f=11$, we have
$P(S|f=11)=\eta p_{11}$. For the function $01$, on the other hand, we succeed if there is no
detection, which has
probability
$(1-p_{01})+(1-\eta) p_{01}$. The first term in the sum
corresponds to the case where the photon goes to a point away from the detector, while in the second
the photon hit the detector but the detector fails. Of course, we have also
$P(S|f=10)=(1-p_{10})+(1-\eta)
p_{10}$. Since the experiment is designed in a such a way that $p_{00}\approx p_{11}\equiv p_{c}$
and $p_{01}\approx p_{10}\equiv p_{b}$, we have finally:
\begin{equation}
 P(S)=\frac{1}{2}[1+\eta(p_{c}-p_{b})].
\end{equation}

Of course, for any detector width, the probability is just $1/2$ for $\eta=0$, since we do not
gain any information about the function. It then grows linearly with $\eta$ and, for the detector
width we have used (100$\mu$m), it goes to a maximum of 0.55.

\begin{figure}[ht!]
    \begin{center}
        \includegraphics[width=9cm]{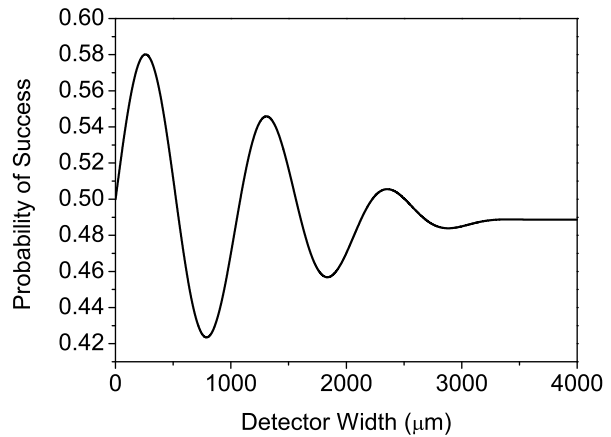}
        \end{center}
        \caption{Probability of success when the signal detector 
        size is changed (width of the single slit in front of the detector is varied ). 
        The photon detectors are assumed to be fixed at the positions $x_{i} = x_{s} = 0$, and it is
assumed the detection of the idler photon. Here the detectors are considered perfect, and we
consider the average  probability of both detection and no-detection events. It is  assumed that we
always bet in a constant (balanced) function when there is (no) detection. The optimal size for the
detector is 260$\mu$m, with a success probability of 0.58.}
    \label{fig:Fig5}
\end{figure}

We could also vary the detector's size by changing the width of the
slit
placed in front of it,
to maximize the right function choice. Figure~\ref{fig:Fig5} shows the
success probability when the slit size is changed, and the detector is
considered perfect. The best detector slit size is 260 $\mu m$ and the corresponding success
probability is 0.58. For certain values the
probability of detecting photons can be larger for balanced functions than for constant ones, so we
would infer the function wrongly more often than correctly. Of course, we would then have to bet in
the opposite way: balanced if we detect a photon, constant if we do not detect anything. For a
detector covering the whole plane we recover the classical fifty-fifty guess since, again, we do
not get any information about $f$.

We note finally that in this setting we have an asymmetry between the betting confidence on
constant and balanced functions: For small detector sizes, a detection implies a constant
function with high
probability, while no detection implies a balanced one just with
moderate probability. Indeed, from Bayes' formula we can compute:
\begin{eqnarray}
 P(f=\text{constant}|\text{detection})=\frac{p_{c}}{p_{c}+p_{b}},\\
P(f=\text{balanced}|\text{no-detection})=\frac{1-\eta p_{b}}{2-\eta(p_{c}+p_{b})}.
\end{eqnarray}

For instance, for a very small detector we have $p_{b}\ll p_{c}\ll 1$ so $
P(f=\text{constant}|\text{detection})\approx 1$ while $
P(f=\text{balanced}|\text{no-detection})\approx 1/2$.
 This is due to our choice of the
detector position. Choosing a spot on the Fourier plane corresponding to the
state $\ket{-}$ would invert this asymmetry. If a CCD was used
instead we would have, on average, the same confidence for betting
on both types of functions, but there would still be inconclusive events. If the signal
photon were detected by the CCD on a position corresponding to the state preparation, say,
$\frac{1}{\sqrt{2}}(\ket{0}+i\ket{1})$, we could not infer the type
of function we had. For instance, a detection at position 
$0.11 mm$, {\it{i.e.}}, where the two interference patterns cross in Fig.~\ref{fig:Fig2},
corresponds to such a situation.

\vspace{0.0cm}

\section{Conclusions}

In this work we implement the minimal Deutsch algorithm version with one qubit, using spontaneous
parametric down conversion and idler photodetection as a heralded source of one photon and a spatial
light modulator as the key part of the quantum oracle. A double-slit is used to encode a qubit in
the photonic transversal spatial modes and the state is manipulated using the SLM. The experimental
setup is able to implement all possible one bit constant and balanced functions easily. Furthermore,
we discuss the improvement of this specific quantum experimental
implementation, when compared to the analogous classical algorithm. This is the first
quantum algorithm implemented using a SLM,
 opens the possibility to implement more complex algorithms.
For example, by changing the double-slit to an eight-slit interferometer \cite{MOMENTA}, one can
also implement the analogous minimal version of Deutsch-Jozsa algorithm for three-bit functions
$f:\{0,1\}^3\rightarrow\{0,1\}$.

\begin{acknowledgments}
 This work is part of the Brazilian National Institute for Science and
Technology for Quantum Information and was supported by the Brazilian agencies CNPq, CAPES, and
FAPEMIG. We acknowledge the \textit{EnLight} group and P. L. de Assis, for very useful discussions.

\end{acknowledgments}


\begin{thebibliography}{10}

\bibitem{DEUTSCH}
D.~Deutsch.
\newblock {\em Proc. R. Soc. Lond. A}, 400:97--117, 1985.

\bibitem{BETTER}
D.~Collins, K.~W. Kim, and W.~C. Holton.
\newblock {\em Phys. Rev. A}, 58:R1633, 1998.

\bibitem{NMR1}
I.~L. Chuang, L.~M.~K. Vandersypen, X.~Zhou, D.~W. Leung, and S.~Lloyd.
\newblock {\em Nature}, 393:143--146, 1998.

\bibitem{NMR2}
L.~M.~K. Vandersypen, M.~Steffen, G.~Breyta, C.~S. Yannoni, M.~H. Sherwood, and
  I.~L. Chuang.
\newblock {\em Nature}, 414:883--887, 2001.

\bibitem{ION1}
S.~Gulde, M.~Riebe, G.~P.~T. Lancaster, C.~Becher, J.~Eschner, H.~Haffner,
  F.~Schmidt-Kaler, I.~L. Chuang, and R.~Blatt.
\newblock {\em Nature}, 410:44--50, 2003.

\bibitem{CAV}
V.~Giovannetti, D.~Vitali, and P.~Tombesi.
\newblock {\em Optics and Spectroscopy}, 91:423--428, 2001.

\bibitem{JOSEPH}
J.~Siewert and R.~Fazio.
\newblock {\em Journal of Modern Optics}, 49:1245--1254, 2002.

\bibitem{PHOTON1}
P.~Walther, K.~J. Resch, T.~Rudolph, E.~Schenck, H.~Weinfurter, V.~Vedral,
  M.~Aspelmeyer, and A.~Zeilinger.
\newblock {\em Nature}, 434:169--176, 2005.

\bibitem{PHOTON2}
G.~Vallone, G.~Donati, N.~Bruno, A.~Chiuri, and P.~Mataloni.
\newblock {\em Phys. Rev. A}, 81:R050302, 2010.

\bibitem{DEUTSCH1992rapid}
D.~Deutsch and R.~Jozsa.
\newblock {\em Proceedings: Mathematical and Physical Sciences},
  439(1907):553--558, 1992.

\bibitem{OLIVEIRA2005implementing}
A.~N. Oliveira, S.~P. Walborn, and C.H. Monken.
\newblock {\em Journal of Optics B: Quantum and Semiclassical Optics}, 7:288,
  2005.

\bibitem{BIANUCCI2004single}
P.~Bianucci, A.~Muller, C.K. Shih, Q.Q. Wang, Q.K. Xue, and C.~Piermaroc.
\newblock In {\em International Quantum Electronics Conference}. Optical
  Society of America, 2004.

\bibitem{SPDC}
Z.~Y. Ou and L.~Mandel.
\newblock {\em Phys. Rev. Lett.}, 61:50--53, 1988.

\bibitem{POLARIZATION}
Y.~H. Shih and C.~O. Alley.
\newblock {\em Phys. Rev. Lett.}, 61:2921--2924, 1988.

\bibitem{MOMENTA}
L.~Neves, G.~Lima, J.~G.~Aguirre G\'{o}mez, C.~H. Monken, C.~Saavedra, and
  S.~P\'{a}dua.
\newblock {\em Phys. Rev. Lett.}, 94:100501, 2005.

\bibitem{LONGITUDINAL1}
J.~G. Rarity and P.~R. Tapster.
\newblock {\em Phys. Rev. Lett.}, 64:2495--2498, 1990.

\bibitem{LONGITUDINAL2}
A.~Rossi, G.~Vallone, A.~Chiuri, F.~De Martini, and P.~Mataloni.
\newblock {\em Phys. Rev. Lett.}, 102:153902, 2009.

\bibitem{ORBITAL1}
A.~Mair, A.~Vaziri, G.~Weihs, and A.~Zeilinger.
\newblock {\em Nature}, 412:313--316, 2001.

\bibitem{ORBITAL2}
N.~K. Langford, R.~B. Dalton, M.~D. Harvey, J.~L. O'Brien, G.~J. Pryde,
  A.~Gilchrist, S.~D. Bartlett, and A.~G. White.
\newblock {\em Phys. Rev. Lett.}, 93:053601, 2004.

\bibitem{SLM1}
G.~Lima, A.~Vargas, L.~Neves, R.~Guzm{\'a}n, and C.~Saavedra.
\newblock {\em Optics Express}, 17(13):10688--10696, 2009.

\bibitem{SLM2}
S.~Cialdi, D.~Brivio, and M.~G.~A. Paris.
\newblock {\em Phys. Rev. A}, 81:042322, 2010.

\bibitem{SLM3}
W.~M. Pimenta, B.~Marques, M.~A.D. Carvalho, M.~R. Barros, J.~G. Fonseca,
  J.~Ferraz, M.~Terra Cunha, and S~P\'{a}dua.
\newblock {\em Optics Express}, 18:24423--24433, 2010.

\bibitem{SLM4}
J.~Leach, B.~Jack, J.~Romero, M.~Ritsch-Marte, R.~W. Boyd, A.~K. Jha, S.~M.
  Barnett, S.~Franke-Arnold, and M.~J. Padgett.
\newblock {\em Optics Express}, 17:8287--8293, 2009.

\bibitem{SIMULATION}
G. Puentes, C.~La Mela, S. Ledesma, C. Iemmi, J.~P.
  Paz, and M. Saraceno.
\newblock {\em Phys. Rev. A}, 69:042319, 2004.

\bibitem{LEONARDO}
L.~Neves, S.~P\'{a}dua, and C.~Saavedra.
\newblock {\em Phys. Rev. A}, 69:042305, 2004.

\bibitem{EMARANHANDO}
L.~Neves, G.~Lima, E.~J.~S. Fonseca, L.~Davidovich, and S.~P{\'a}dua.
\newblock {\em Physical Review A}, 76(3):032314, 2007.

\bibitem{MORENO}
I.~Moreno, P.~Vel\'asquez, C.~R. Fern\'andez-Pousa, and M.~M.
  S\'anchez-L\'opes.
\newblock {\em J. App. Phys.}, 94:3697, 2003.

\end{thebibliography}
\end{document}